# Competing spin modulations in a magnetically frustrated semimetal EuCuSb

Hidefumi Takahashi[1,2], Kai Aono[2], Yusuke Nambu[3], Ryoji Kiyanagi[4], Takuya Nomoto[2], Masato Sakano[2], Kyoko Ishizaka[2,5], Ryotaro Arita[2,5] and Shintaro Ishiwata[1,2]

[1]*Division of Materials Physics and Center for Spintronics Research Network (CSRN), Graduate School of Engineering Science, Osaka University, Osaka 560-8531, Japan*

[2]*Department of Applied Physics, The University of Tokyo, Tokyo 113-8656, Japan*

[3]*Institute for Materials Research, Tohoku University, Sendai, Miyagi 980-8577, Japan*

[4]*J-PARC Center, Japan Atomic Energy Agency, Tokai, Naka, Ibaraki 319-1195, Japan*

[5]*RIKEN Center for Emergent Matter Science (CEMS), Wako, Saitama 351-0198, Japan*

The competing magnetic ground states of the itinerant magnet EuCuSb, which has a hexagonal layered structure, were studied via magnetization, resistivity, and neutron diffraction measurements on single-crystal samples. EuCuSb has a three-dimensional semimetallic band structure as confirmed by band calculation and angle-resolved photoelectron spectroscopy, consistent with the nearly isotropic metallic conductivity in the paramagnetic state. However, below the antiferromagnetic transition temperature of $T_{N1}$ (8.5 K), the resistivity, especially along the hexagonal axis, increases significantly. This implies the emergence of anisotropic magnetic ordering coupled to the conducting electrons. Neutron diffraction measurements show that the Eu spins, which order ferromagnetically within each layer, are collinearly modulated (up-up-down-down) along the hexagonal axis below $T_{N1}$, followed by the partial emergence of helical spin modulation below $T_{N2}$ (6 K). Based on the observation of anomalous magnetoresistance with hysteretic behavior, we discuss the competing nature of the ground state inherent in a frustrated Heisenberg-like spin system with a centrosymmetric structure.

Magnetic materials with spin frustration often host competing magnetic ground states with unique spin structures.[1–4] The presence of itinerant electrons coupled to frustrated spins results in unusual magnetotransport phenomena such as the unconventional anomalous Hall effect in noncollinear magnets[5–8] and multi-step magnetoresistance in collinear magnets with various spin modulations.[9,10] In recent decades, topologically nontrivial emergent helical spin structures such as skyrmion lattices and hedgehog-antihedgehog lattices have been extensively studied in frustrated magnets[11–14], which hold promise for novel spintronic functionalities. These helimagnetic phases are typically stabilized in non-centrosymmetric systems, in which the Dzyaloshinskii-Moriya interaction (DMI) causes a spiral spin modulation with a fixed spin chirality.[15,16] On the other hand, recent theoretical works suggest that centrosymmetric systems can potentially exhibit topological spin structures through geometric frustration or multiple long-range interactions beyond the conventional Ruderman-Kittel-Kasuya-Yoshida (RKKY) interaction.[17–23] Because the highly symmetric lattices without DMI are highly symmetric, novel topological magnetic phases such as an antiskyrmion phase and a high-topological number skyrmion phase have been theoretically predicted in centrosymmetric frustrated magnets.

Recently, the itinerant magnet Gd$_2$PdSi$_3$, which has a centrosymmetric space group, was reported to host a skyrmion lattice phase with a gigantic topological Hall effect.[23,24] . Here, we focus on EuCuSb, which is a member of the family of hexagonal ABC-type compounds as well as Gd$_2$PdSi$_3$ (see Fig. 1(a)).[25,26] The Eu and Cu/Sb in EuCuSb form a two-dimensional triangular lattice and honeycomb lattice, respectively. It should be noted here that EuCuSb is free from local inversion symmetry breaking with respect to the two adjacent magnetic ions in the two-dimensional triangular lattice, unlike Gd$_2$PdSi$_3$, where the local inversion symmetry is broken by the Pd/Si ordering.[27] Thus, EuCuSb has the potential to show novel spin textures other than the skyrmion lattice that reflect the absence of DMI.

In this Letter, we report the electronic structure and competing magnetic ground states of the itinerant magnet EuCuSb, which has a centrosymmetric hexagonal structure. The first-principles calculations reveal a semimetallic band structure with nearly three-dimensional Fermi surfaces, as also partly confirmed by angle-resolved photoelectron spectroscopy (ARPES). Through magnetic and resistivity measurements, we found two successive magnetic transitions at $T_{N1}$ = 8.5 K and $T_{N2}$ = 6 K, and an unusual magnetoresistance that indicates phase competition. Neutron diffraction measurements reveal the presence of competing magnetic phases with spin modulation along the hexagonal axis at low temperatures, and suggested the coexistence of commensurate collinear and incommensurate helical spin structures below $T_{N2}$. We discuss the origin of these unique magnetic phase transitions and the possible emergence of nontrivial spin structures under a magnetic field in the magnetically frustrated semimetal EuCuSb.

Single crystals of EuCuSb with typical dimensions of 1×1×5 mm$^3$ (see Fig. 1(b)) were grown by the flux method with bismuth, as described in Ref. [28]. The structural parameters at room temperature were refined using a Rigaku XtaLAB-mini II diffractometer with graphite monochromated Mo *Kα* radiation (space group: *P*6$_3$/mmc, *a* = 4.513(4) Å, *c* = 8.546(3) Å). The magnetic and transport properties were measured using a magnetic property measurement system (Quantum Design, Inc.). The ARPES measurements were performed by using DA30 electron analyzer (Scienta Omicron) and 6.42 eV laser light source at Department of Applied Physics, the University of Tokyo. The crystal was cleaved at (100) plane under low-temperature (~15 K) and ultra-high vacuum condition (~1×10$^{-10}$ Torr). The energy resolution was set to 1 meV. Though the work function can not be determined experimentally to date because of low photon energy, we temporarily adopted the value of 4.45 eV. The relativistic electronic structure of EuCuSb was calculated by density functional theory (DFT) using the VASP code with open-core Eu-4f electrons. The projector augmented wave (PAW) method was used to account for the core electrons[29,30]. We used the Perdew-Burke-Ernzerhof (PBE) exchange-correlation functional. The Brillouin zone was sampled by a 12×12×12 *k*-mesh for the PWscf calculation, and a 20×20×20 *k*-mesh for the Fermi surface visualization by the FermiSurfer package.[31] Single-crystal neutron diffraction measurements on EuCuSb were performed using the time-of-flight (TOF) single-crystal neutron diffractometer SENJU installed at J-PARC.[32]

Let us first discuss the basic electronic structure of EuCuSb. The Fermi surfaces of EuCuSb obtained by the first-principles band calculation are shown in the 3-dimensional Brillouin zone in Fig. 1(c).

They are characterized by the hole and electron pockets with ellipsoidal shapes around the Γ and M points, respectively, thus indicating the semi-metallic electronic structure. To experimentally confirm this, we performed the 6.42 eV laser-based ARPES on EuCuSb (100) plane at 15 K, and successfully obtained the Fermi surfaces in the momentum plane spanned by Γ-A and Γ-M (Fig. 1(e)). We can clearly see the two elliptical Fermi surfaces consisting of hole-like bands that surround the Γ point, with their longer axes along the Γ-A direction. They are quite well reproduced in Fig. 1(d), suggesting the validity of the band calculation.

The temperature dependence of the magnetic susceptibility χ for external fields $H//c$ (red) and $H \perp c$ (blue) is shown in Fig. 2(a). The data collected on cooling at $H = 0.1$ T and on warming with the same field after zero-field cooling exhibited no bifurcation between the field-cooled and zero-field-cooled protocols. A Curie-Weiss-like behavior was observed under magnetic fields $H//c$ and $H \perp c$, followed by two successive antiferromagnetic transitions at $T_{N1}$ (= 8.5 K) and $T_{N2}$ (= 6 K). The two successive transitions were also observed in a previous study with polycrystalline samples.[25] A Curie-Weiss fit to the χ-T curve for $H//c$ gives a Weiss temperature of $\theta_w = 4$ K, which reflects the subtle predominance of the ferromagnetic interaction. The estimated effective moment $p_{eff} = 7.7$ $\mu_B$/f.u. is close to the theoretical value ($p_{eff} = 7.94$ $\mu_B$/f.u.) for $Eu^{2+}$ ions with $S = 7/2$. The antiferromagnetic order and positive $\theta_w$ suggest the competing nature of the magnetic interactions between the $Eu^{2+}$ ions.

Figure 2(b) displays the temperature dependence of the out-of-plane ($I//c$) and in-plane ($I \perp c$) resistivities. The out-of-plane resistivity was measured in two ways called the *untrained* ($\rho_{ut}$) and *trained* ($\rho_t$) states, as described in Ref.[33]. $\rho_{ut}$ was measured in the cooling process under a zero magnetic field (solid red line). On the other hand, $\rho_t$ was measured using the following procedure: (i) a magnetic field ($H \perp c$) of 7 T was first applied at 20K ($> T_{N1}$), (ii) the sample was then cooled down to 2 K under 7 T, (iii) the magnetic field was next set to zero, and the (iv) resistivity was finally measured in the warming process (dashed red line). The out-of-plane untrained resistivity $\rho_{ut}$ exhibits metallic behavior from 300 to 50 K and then sharply increases with decreasing temperature below $T_{N1}$, followed by a subtle suppression below $T_{N2}$. Interestingly, $\rho_t$ further increases upon cooling below $T_{N2}$, unlike $\rho_{ut}$. This result suggests that the application of the magnetic field changed the magnetic domain structure below $T_{N2}$ even in the absence of spontaneous ferromagnetism, as has been reported for pyrochlore iridates.[33,34] The temperature dependence of the in-plane (untrained) resistivity is similar to that of the out-of-plane $\rho_{ut}$, while the anisotropy of ρ is enhanced below $T_{N2}$. The nearly isotropic temperature dependence of ρ, especially above $T_{N1}$, is consistent with the three-dimensional ellipsoidal Fermi surface revealed by the ARPES measurement at 15 K.

To clarify the influence of the magnetic field on the transport properties, we measured the magnetization and resistivity (out-of-plane) as functions of $H$ at 2 K, as shown in Fig. 3. For $H//c$, the magnetization varies linearly with the magnetic field in the low-field region and saturates around 4.3 T (= $H^c_{c1}$) without visible hysteresis (Fig. 3(a)). The saturated moment of approximately 6.2 $\mu_B$/f.u. is slightly smaller than the theoretical saturated moment of $Eu^{2+}$ (7 $\mu_B$/f.u.), implying the presence of small concentrations of $Eu^{3+}$ ions. On the other hand, the magnetization curve for $H \perp c$ shows a tiny kink around 1.3 T (= $H^{ab}_{c1}$) with hysteretic behavior(Fig. 3(b)), which implies a spin-flop transition at $H^{ab}_{c1}$. The saturated field ($H^{ab}_{c2} \sim 4.0$ T) and moment (~ 6.5 $\mu_B$/f.u.) are almost the same as those under $H//c$, suggesting that the local moment has a Heisenberg-like character with a small easy plane-type anisotropy at 2 K.

The magnetoresistance shown in Figs. 3(c) and 3(d) were measured using the following procedure [33]: the sample was first cooled down to 2 K under zero magnetic field, and then the resistivity was measured by sweeping the magnetic field in the follow sequence: (1) from 0 T to 7 T, (2) from 7 T to 0 T, and (3) from 0 T to 7 T. Upon increasing $H$ along $c$, the longitudinal magnetoresistance decreases with a slight deviation between process (1) and processes (2) and (3), as shown in Fig. 3(c). In stark contrast to this behavior, the transverse magnetoresistance ($H \perp c$) presented in Fig. 3(d) shows a remarkable deviation between processes (1), (2), and (3). The resistivity is as small as 0.9 mΩcm at $H = 0$ T and undergoes an irreversible jump around $H^{ab}_{c1}$ in process (1). As the field decreases from 7 T [process (2)], the resistivity monotonically increases and forms kinks around 1.5 T and $H^{ab}_{c1}$, and then drops below 0.7 T, reaching a maximum value around 0.5 T. It should be noted that the resistivity in process (2) no longer tracks the resistivity in process (1) and is significantly larger than that of the untrained state at $H = 0$ T. The resistivity in process (3) shows a similar profile to that in process (1) with a smaller jump around 1.5 T. While such a hysteretic magnetoresistance is reminiscent of the field-induced alignment of the magnetic domains seen in ferromagnetic metals, the positive magnetoresistance seen in the present system is opposite to what is expected for a ferromagnetic metal.[35] On the other hand, the positive magnetoresistance accompanied by the domain alignment is reminiscent of domain wall conductance observed in iridium pyrochlores.[32]

To gain a microscopic understanding of the magnetism, we performed neutron diffraction measurements using a large single-crystal sample (2×2×4 mm³). Figures 4(a), 4(b), and 4(c) display the neutron diffraction profiles on the ($h\ 0\ l$) reciprocal lattice plane at 11 K ($> T_{N1}$), 7 K ($T_{N2} < T < T_{N1}$), and 3 K ($< T_{N2}$), respectively. At 11 K, nuclear Bragg reflections with the $P6_3/mmc$ symmetry were observed at the integer $h$ and $l$ positions. The large absorption cross section of Eu (4530 barn) necessitated a longer counting time to obtain clear reflections. Therefore, the data are contaminated by background noise, where the ring-like reflections originated from the aluminum metal used for the cryostat and sample holder.

Below $T_{N1}$, additional magnetic reflections are observed (yellow arrows in Fig. 4(b)). These can be accounted for by the commensurate magnetic wave vector $\boldsymbol{q}_{m1} = (0, 0, 1/2)$, which implies a doubled magnetic unit cell along the $c$-axis ($a \times b \times 2c$). The absence of magnetic reflections along the $h = 0$ line indicates that the Eu magnetic moments point along the $c$-axis, which is reflected in the polarization factor in magnetic neutron scattering.[36] We employed group theoretical analysis to identify the magnetic structure, and found that only the commensurate and collinear spin structure with the up-up-down-down (uudd) arrangement (see Fig. 4(e)) is allowed by symmetry.

Interestingly, below $T_{N2}$, other magnetic reflections (denoted by blue arrows in Fig. 4(c)) emerge with a separate incommensurate magnetic wave vector $\boldsymbol{q}_{m2} = (0, 0, 0.3)$ in addition to the $q_{m1}$ phase. The appearance of peaks from both $q_{m1}$ and $q_{m2}$ along the $h = 0$ line indicates that the moments possess $ab$-plane components below $T_{N2}$. Given that the clear hysteresis in the magnetotransport and the deviation of $\rho_{ut}$ and $\rho_t$, the magnetic ground state is likely composed of the spin-reoriented uudd and incommensurate ($\boldsymbol{q}_{m2}$) magnetic structures with large domains. $\boldsymbol{q}_{m2}$ allows for two candidate magnetic structures, namely, the sinusoidal and helical structures, both of which have in-plane spin components. We performed the absorption correction on the obtained data and carried out a magnetic structure analysis. Although the absorption correction is not perfect because it was not possible to obtain an accurate estimate of the sample dimensions, the analysis appears to favor the helical structure with a lower reliability factor. Given that $Eu^{2+}$ spins usually have a small magnetic anisotropy that reflects the quenched orbital moment evidenced by the magnetization (Fig. 3) measurements and the absence of uniaxial structural anisotropy in the *ab* plane, the helical spin structure is likely to be realized as a characteristic of the Heisenberg system[37]. Figures 4(d), 4(e), and 4(f) depict schematic illustrations of the temperature variation of the spin structures. The uudd phase, in which spins are aligned along the *c*-axis, emerges below $T_{N1}$. These spins tilt in the in-plane direction across $T_{N2}$, and coexist with another incommensurate helimagnetic phase; however, precise polarized neutron measurements are needed to quantitatively determine the ratio of the in-plane and out-of-plane spin components.

Here, we discuss the origin of the uudd phase and the unusual coexistence of the commensurate uudd and incommensurate helimagnetic phases. One of the simplest models relevant for EuCuSb is the classical one-dimensional Heisenberg spin chain with competing nearest-neighbor ($J_1$) and next-nearest-neighbor ($J_2$) interactions and nearest-neighbor biquadratic interactions ($J_b$)[38]. In the ground state phase diagram of this model, the uudd and helimagnetic states are stabilized and compete with each other when $J_2/|J_1|$ is larger than 0.5. The uudd state is favored by the biquadratic interaction $J_b$. Because the spin modulation in EuCuSb propagates perpendicular to the *ab* plane on which the Eu spins are aligned ferromagnetically, this compound can be regarded as a one-dimensional spin chain system with competing exchange interactions along the *c* axis, as described in the above model. Given that the ground state of this compound is located on the phase boundary between the uudd and helimagnetic phases, the subtle chemical disorder may result in the coexistence of these phases in a multi-domain state. However, it should be noted that this model applies for localized spin systems. Another possibility is the oscillatory RKKY interaction through the itinerant electrons, which can be a source of the complicated spin structure in Ising-like spin systems [39]. However, because the $Eu^{2+}$ spin is typically isotropic, an additional anisotropic interaction is needed for stabilizing the uudd phase in EuCuSb. While we should also consider other possible mechanisms such as spin-density-wave type interactions associated with electronic structural instability[40] and inter-orbital frustration in rare-earth-based compounds[41], more detailed investigations on the electronic and magnetic structures of EuCuSb are necessary to discuss these possibilities.

Finally, we comment on the possibility of a nontrivial topological magnetic structure in EuCuSb. Recent theoretical works have proposed that centrosymmetric frustrated magnets with rotational symmetry have the potential for exhibiting novel topological spin textures characterized by multiple-*q* spin spirals[17]. In fact, novel topological magnetic phases have recently been discovered in itinerant frustrated magnets with inversion symmetry.[22,23]. For EuCuSb, such a topological magnetic phase may be induced by the application of the magnetic field. Some anomalies implying magnetic phase transitions have been observed in the magnetoresistance under both $H//c$ (denoted by an open triangle) and $H \perp c$ (denoted by $H^{ab}_{c1}$ and open triangles) at low temperatures.

In conclusion, we have investigated the magnetic and transport properties of the itinerant magnet EuCuSb and performed single-crystal neutron diffraction on the material. Two successive magnetic transitions ($T_{N1}$ = 8.5 K, $T_{N2}$ = 6 K) were identified from the magnetic susceptibility and resistivity measurements. The neutron diffraction measurements revealed a commensurate and collinear spin ordering (uudd phase) below $T_{N1}$, followed by the emergence of the incommensurate helimagnetic phase coexisting with the uudd phase with a spin reorientation transition below $T_{N2}$. These results suggest that EuCuSb is a layered frustrated magnet, which can be regarded as a one-dimensional Heisenberg-like spin chain with competing exchange interactions. EuCuSb may potentially exhibit novel topological magnetic phases under a magnetic field.

**Acknowledgements**

The authors thank H. Kawamura, K. Aoyama, K. Mitsumoto, Y. Tanaka, and Y. Narumi for fruitful discussions, and M. Avdeev for preliminary powder neutron diffraction measurements. This study was supported in part by KAKENHI (Grant No. 17H01195, 17H06137, 19H02424, 19K14652, and 20K03802), JST CREST (Grant No. JPMJCR16F1), and the Asahi Glass Foundation.

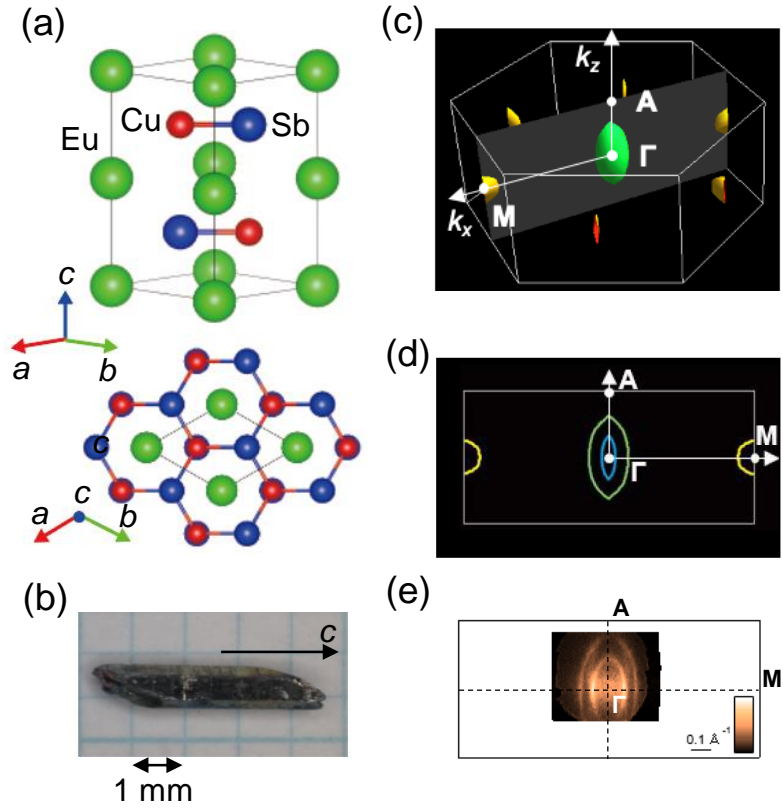

FIG. 1. (a) Crystal structure of EuCuSb. (b) Photograph of the single-crystal sample. (c) Fermi surfaces obtained by the band calculation in 3-dimensional Brillouin zone. (d) Cross-sectional view of the Fermi surfaces in the momentum plane including the G, A, and M points. (e) The image of Fermi surfaces obtained by 6.42 eV laser-based ARPES on (100) plane EuCuSb recorded at 15 K. The color-scale indicates the ARPES intensity.

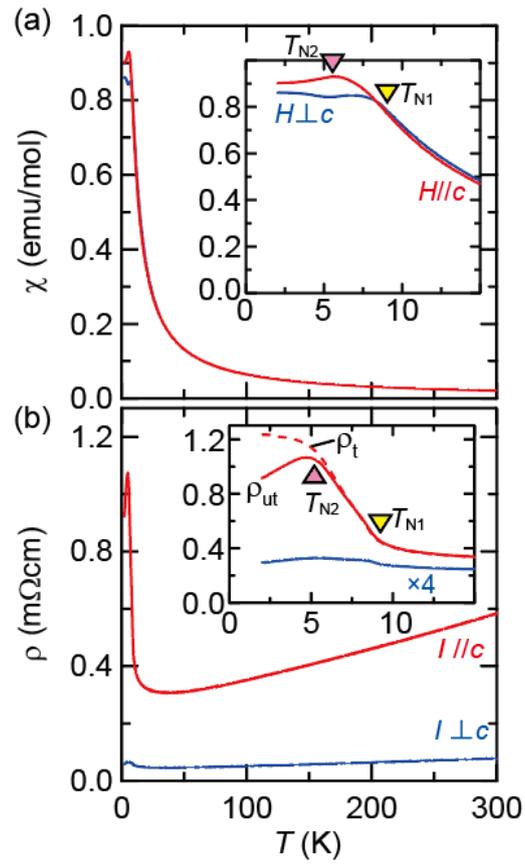

FIG. 2. Temperature dependence of (a) the magnetic susceptibility under $H//c$ and $H\perp c$ and (b) the out-of-plane ($I//c$) and in-plane ($I\perp c$) resistivity. The inset to (b) shows the resistivity of the untrained state $\rho_{ut}$ and that of the trained state $\rho_t$.

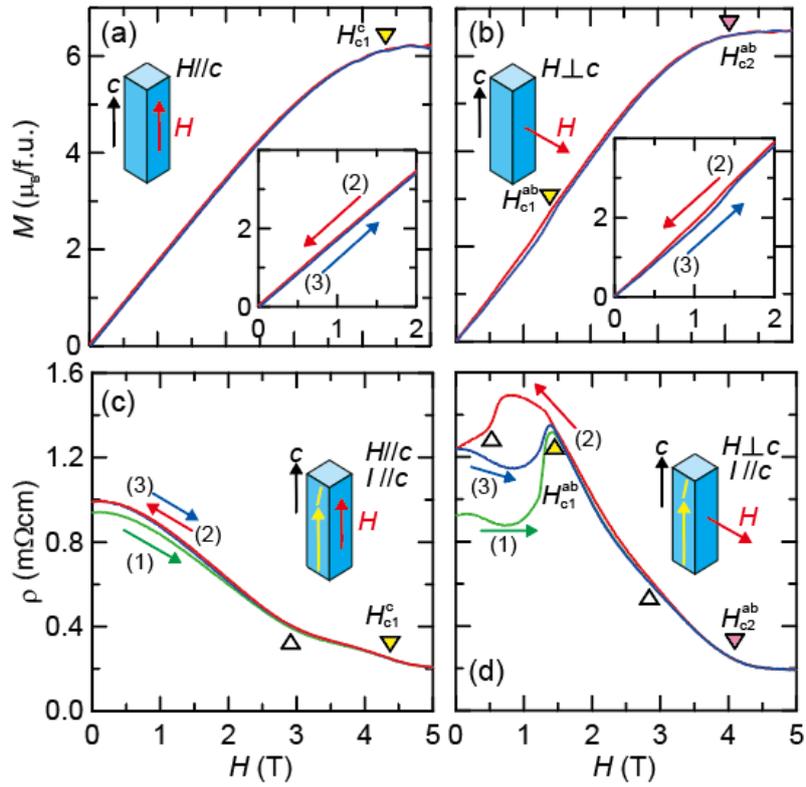

FIG. 3. Magnetic field dependence of magnetization under (a) $H//c$ and (b) $H\perp c$ at 2 K. Magnetic field dependence of resistivity under (c) longitudinal ($H//c$) and (d) transverse ($H\perp c$) fields at 2 K. The green curve denotes the virgin curve [process (1)], the red curve the sweeping process (2), and the blue curve the process (3).

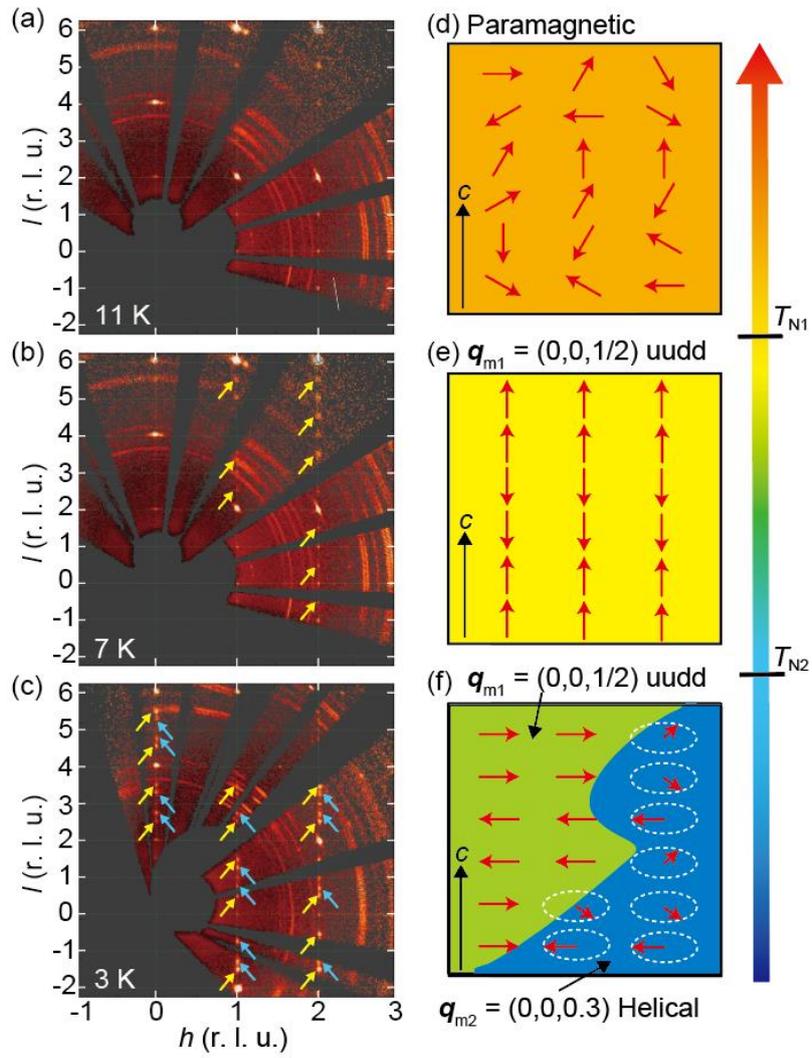

FIG. 4. Neutron diffraction intensity distributions of EuCuSb on the ($h\ 0\ l$) reciprocal lattice plane at (a) 11 K ($> T_{N1}$), (b) 7 K ($T_{N2} < T < T_{N1}$), and (c) 3 K ($< T_{N2}$). The blue and yellow arrows in (b) and (c) indicate the magnetic reflections from the Eu sublattice with the propagation vectors of $\boldsymbol{q}_{m1} = (0, 0, 1/2)$ and $\boldsymbol{q}_{m2} = (0, 0, 0.3)$, respectively. The ring-like intensities correspond to the powder lines which may be due to aluminum in the sample holder. Schematic magnetic structures in the (d) paramagnetic, (e) uudd, and (f) uudd and incommensurate helimagnetic domain phases. Because the ratio of the $ab$ and $c$ components of the spin moment is ambiguous, the spin direction in (f) is the projected component on the $ab$ plane.